*Data Descriptor*

# A Large-Scale Dataset of Twitter Chatter about Online Learning during the Current COVID-19 Omicron Wave

**Nirmalya Thakur**

Department of Electrical Engineering and Computer Science, University of Cincinnati, Cincinnati, OH 45221-0030, U.S.A.; thakurna@mail.uc.edu

**Abstract:** The COVID-19 Omicron variant, reported to be the most immune evasive variant of COVID-19, is resulting in a surge of COVID-19 cases globally. This has caused schools, colleges, and universities in different parts of the world to transition to online learning. As a result, social media platforms such as Twitter are seeing an increase in conversations related to online learning in the form of tweets. Mining such tweets to develop a dataset can serve as a data resource for different applications and use-cases related to the analysis of interest, views, opinions, perspectives, attitudes, and feedback towards online learning during the current surge of COVID-19 cases caused by the Omicron variant. Therefore, this work presents a large-scale open-access Twitter dataset of conversations about online learning from different parts of the world since the first detected case of the COVID-19 Omicron variant in November 2021. The dataset is compliant with the privacy policy, developer agreement, and guidelines for content redistribution of Twitter, as well as with the FAIR principles (Findability, Accessibility, Interoperability, and Reusability) principles for scientific data management. The paper also briefly outlines some potential applications in the fields of Big Data, Data Mining, Natural Language Processing, and their related disciplines, with a specific focus on online learning during this Omicron wave that may be studied, explored, and investigated by using this dataset.

## 1. Introduction

The first cases of the COVID-19 pandemic, caused by the SARS-CoV-2 virus, were recorded in a seafood market in Wuhan, China, in December 2019 [1]. Since then, the virus has been found in all the countries of the world. At the time of writing this paper, globally, there have been 535,342,382 cases with 6,320,324 deaths [2]. Since the initial cases in China, the SARS-CoV-2 virus has undergone multiple mutations, and as a result, multiple variants have been detected in different parts of the world. Some of these include – Alpha (B.1.1.7), Beta (B.1.351), Gamma (P.1), Delta (B.1.617.2), Epsilon (B.1.427 B.1.429), Eta (B.1.525), Iota (B.1.526), Kappa (B.1.617.1), Zeta (P.2), Mu (B.1.621, B.1.621.1), and Omicron (B.1.1.529, BA.1, BA.1.1, BA.2, BA.3, BA.4 and BA.5) [3]. Out of all these variants, the Omicron variant, first detected on 24th November 2021 from a sample collected on 9th November 2021, was classified as a Variant of Concern (VOC) by World Health Organization (WHO) on 26th November 2021 [4]. The Omicron variant has a spike protein that contains 30 mutations [5]. It has been reported to be the most immune evasive variant of COVID-19 and to present very strong resistance against antibody-based or plasma-based treatments [6]. According to WHO, the new cases due to this Omicron variant have been "off the charts" and are setting new records in terms of COVID-19 cases all over the world [7]. The Omicron variant currently accounts for 86% of the COVID-19 cases worldwide [8], and some of the countries that have recorded the most cases due to the SARS-CoV-2 Omicron variant include – the United Kingdom (1,138,814 cases), USA (945,470 cases), Germany (245,120 cases), Denmark (218,106 cases), France (110,959 cases), Canada (92,341 cases), Japan (71,056 cases), India (56,125 cases), Australia (46,576 cases), Sweden (43,400 cases), Israel (39,908 cases), Poland (33,436 cases), and Brazil (32,880 cases) [9].

Since the beginning of the pandemic, many countries in the world, such as – India [10], United States [11], United Kingdom [12], Spain [13], Greece [14], Italy [15], Austria [16], Nigeria [17], China [18], New Zealand [19], Ireland [20], Germany [21], South Africa [22], Australia [23], France, [24], Norway [25], and several more [26], went on a complete lockdown with work from home and remote work guidelines that affected a multitude of industries and sectors. Out of all these sectors that were impacted by the nationwide lockdowns and the associated guidelines in different parts of the world, the education sector was an important one. On a global scale, universities, colleges, and schools had to switch to online education, which required its faculty, administrators, staff, and students to get familiarized with online learning and the associated tools and platforms that were necessary for this new norm of education. Due to the worldwide adoption and familiarization with various forms of tools, platforms, software, and hardware necessary for online education, the online education market is rapidly booming and is expected to reach more than USD 350 billion by 2025 [29]. Online learning may be broadly defined as *"learning experiences in synchronous or asynchronous environments using different devices (e.g., mobile phones, laptops, etc.) with internet access. In these environments, students can be anywhere (independent) to learn and interact with instructors and other students"* [27]. Online learning has a range of synonyms, and some of the most commonly used synonyms include remote education, online education, virtual education, remote learning, e-learning, distance education, virtual learning, asynchronous learning, and blended learning [27].

On a global scale, more than 43,518,726 students were affected due to in-person school closures due to COVID-19 [28]. The closing of universities, colleges, and schools was recorded in 188 countries [30], and 90% of the countries reported a switch to one or more forms of online learning [31]. Despite these promising numbers, 31% (463 million) of students in schools (in pre-primary to secondary education) could not adopt online learning either due to lack of technologies, training, or accessibility, and 75% of students who belonged to the poorest households could not switch to the technologies required for online learning [31].

With the advancements in vaccine research and other forms of treatment of COVID-19 towards the later part of 2020 [32-34] and in compliance with the recommendations from various local and national policy-making bodies, different universities, colleges, and schools started to transition to hybrid learning (both online and in-person) as well as completely in-person learning [35]. However, this was associated with several challenges [36], including a surge of COVID-19 cases in students, educators, and staff members, an increase in stress and anxiety in both students and their parents, and the need for allocation of funds by these educational institutions to conduct classes in a socially distant manner, for procurement of hand sanitizers, and disinfectants. Despite these challenges, education continued in both hybrid and in-person forms for a few months. However, due to the recent global surge in COVID-19 cases due to the Omicron variant [7-9], many educational institutions all over the world have transitioned back to online learning since the beginning of 2022, and several are in the process of transitioning to online learning over the next few months [37-42].

The modern-day Internet of Everything lifestyle [43] is characterized by people spending more time on the internet than ever before, with a specific focus on social media platforms. The use of social media platforms has skyrocketed in the recent past [44]. Social media usage characteristics include conversations on diverse topics such as recent issues, global challenges, emerging technologies, news, current events, politics, family, relationships, and career opportunities [45]. Twitter, one such social media platform, used by people of almost all age groups [46,47], has been rapidly gaining popularity in all parts of the world and is currently the second most visited social media platform [48]. At present, there are about 192 million daily active users on Twitter, and approximately 500 million tweets are posted on Twitter every day [49]. Mining of social media conversations, for instance, Tweets, to develop datasets has been of significant interest to the scientific community in the areas of Big Data, Data Mining, and Natural Language Processing, as can be seen from these recent works where relevant Tweets were mined to develop Twitter datasets on 2020 US Presidential Elections [50], 2022 Russia Ukraine war [51], climate change [52], natural hazards [53], European Migration Crisis [54], movies [55], toxic behavior amongst adolescents [56], music [57], civil unrest [58], drug safety [59], and Inflammatory Bowel Disease [60].

In the context of the recent surge of COVID-19 cases due to the Omicron variant and its impact on the education sector, there has been a significant increase in conversations on Twitter related to online learning. Mining such conversations to develop a dataset would serve as a rich data resource for the investigation of different research questions in the fields of Big Data, Data Mining, Data Science, and Natural Language Processing, with a central focus on analyzing tweets related to online learning during this time.

Previous works [61-90] (discussed in Section 2) related to online learning since the outbreak of COVID-19 have focused on analyzing multiple factors related to online learning only in certain geographic regions, mostly by using surveys, and not on a global scale by analyzing conversations from all over the world, such as Tweets. Prior works on the development of Twitter datasets related to COVID-19 have also not focused on mining relevant tweets related to online learning during the ongoing COVID-19 Omicron wave. To address these limitations, this work proposes a dataset of more than 50,000 Tweet IDs (that correspond to the same number of Tweets) about online learning that was posted on Twitter from 9th November 2021 to 13th July 2022, which is publicly available at https://doi.org/10.5281/zenodo.6837118. The earliest date was selected as 9th November 2021, as the Omicron variant was detected for the first time in a sample that was collected on this date. 13th July 2022 was the most recent date at the time of re-submission of this journal paper after the completion of the first round of peer review and the subsequent editorial decision.

The rest of the paper is organized as follows. Section 2 presents an overview of recent works in this field. The methodology that was followed for the development of this dataset is presented in Section 3. Section 4 provides the description of the dataset. Section 5 briefly discusses a few potential applications of this dataset. The conclusion and scope for future work are presented in Section 6, which is followed by references.

## 2. Literature Review

There has been a significant amount of research related to online learning since the global outbreak of COVID-19. The work by Muhammad et al. [61] was a research study that examined the attitudes of Pakistani higher education students towards compulsory digital and distance learning courses during COVID-19. In [62], Rasmitadila et al. presented a study that explored the perceptions of primary school teachers towards online learning during COVID-19. Data were collected through surveys and semi-structured interviews and 67 teachers in primary schools participated in this study. The work by Irawan et al. [63] aimed to identify the impact of student psychology on online learning during the COVID-19 pandemic. The research method used a qualitative research type of phenomenology. The research subjects were 30 students of Mulawarman University, a university in Indonesia, who were interviewed via telephone. The work of Baticulon et al. [64] was to identify barriers to online learning from the perspective of medical students in the Philippines. The authors sent out an electronic survey to the students who participated in this study. The qualitative study presented by Hussein et al. [65] aimed to investigate the attitudes of undergraduate students towards online learning during the first few weeks of the mandatory shift to online learning caused by COVID-19. Students from two general English courses at a university located in the United Arab Emirates were asked to write semi-guided essays and the associated data were analyzed by the authors. The work of Famularsih et al. [66] focused on studying the utilization of online learning applications in English as a Foreign Language (EFL) classrooms. The participants of this study were 35 students from a university in Salatiga, Indonesia. The data was gathered through surveys and semi-structured interviews.

The study by Sutarto et al. [67] focused on understanding the strategies used by teachers of SDIT Rabbi Radhiyya Curup, a school in Indonesia, to increase student's interest and responses to online learning during COVID-19. The data were collected by conducting semi-structured interviews, which were analyzed using the Miles and Huberman model. Almusharraf et al. [68]'s work aimed to evaluate the level of postsecondary student satisfaction with online learning platforms and learning experiences during the COVID-19 pandemic in Saudi Arabia. Quantitative research was carried out in this study by using a survey that was sent out to 283 students enrolled at a higher education institution in Saudi Arabia. These data were analyzed using SPSS. Al-Salman et al. [69] investigated the influence of digital technology, instructional and assessment quality, economic status and psychological state, and course type on Jordanian university students' attitudes towards online learning during the COVID-19 emergency transition to online learning. A total of 4,037 undergraduate students from four universities participated in this study.

The aim of Bolatov et al.'s work [70] was to compare the differences between the mental state of students switching to online learning and the mental state of the students who were still using traditional learning. This study included medical students from Astana Medical University, a university in Kazakhstan. The work by Agormedah et al. [71] explored the responses of students to online learning in higher education in Ghana. The sample size of this study involved 467 students. The findings indicated that majority of the students had a positive response to the transition to online learning. The work of Moawad et al. [72] aimed to identify the academic stressors by analyzing the worries and fears that students at the College of Education in King Saud University, a university in Saudi Arabia, experienced during the time of COVID-19. The results showed that the issue with the highest percentage of stress among students was their uncertainty over the end-of-semester exams and assessments. The work by Khan et al. [73] discussed various digital education methods, approaches, and systems that could be implemented by the education system of Bangladesh during COVID-19. The purpose of the study performed by Catalano et al. [74] was to determine teacher perceptions of students' access and participation in online learning, as well as concerns about educational outcomes among different groups of learners. The work of Kapasia et al. [75] aimed to assess the impact of the nationwide lockdown on account of COVID-19 on undergraduate and postgraduate students in West Bengal, a state in India. The authors conducted an online survey that included 232 students. In [76], Burns et al. performed a conceptual analysis on student wellbeing at universities in the United Kingdom with a specific focus on the psychosocial impact the pandemic had on students. Küsel et al. [77] performed a study to evaluate German university students' readiness for using digital media and online learning in their tertiary education and compared the findings with the results from the same study performed on students in the United States. A total of 72 students from universities in Germany and 176 students from universities in the United States were a part of this study. Darayseh et al. [78] analyzed the impact of COVID-19 on modes of teaching with a specific focus on science education in schools in the United Arab Emirates. Questionnaires were deployed through an online platform, and a total of 62 science teachers participated in this study. Tsekhmister et al. [79] conducted a study to evaluate the effectiveness of virtual reality technology and online teaching systems among medical students of Bogomolets National Medical University, a university in Ukraine. The study was performed using a questionnaire that contained 15 questions with five options to comprehensively evaluate these technologies.

Arsaliev et al.'s work [80] aimed to investigate whether an online format was effective in providing education for ethnocultural competence development. A combination of digital surveys, tests, questionnaires, and online class interviews were used in this study that involved 120 students at Southern Federal University, a university in Russia. Cárdenas-Cruz et al.'s [81] work aimed to facilitate the acquisition of specific transversal skills of undergraduate students at the University of Granada in Spain during the outbreak by means of an integrated online working system. Papouli et al. [82] aimed to explore Greek social work students' views on the use of digital technology during their stay at home due to the coronavirus lockdown. A total of 550 students from different universities in Greece participated in this study. In [83], Parmigiani et al. designed a qualitative study aimed at investigating the factors affecting e-inclusion during COVID-19. A total of 785 teachers at the University of Genoa, a university in Italy, participated in this study. Resch et al. [84] focused on analyzing the effects of COVID-19 on university students' social and academic integration, based on Tinto's integration theory. A total of 640 university students in Austria completed an online survey pertaining to academic and social integration in this study. The purpose of the study by Noah et al. [85] was to examine the impacts of Google classroom as an online learning delivery platform in a secondary school during the COVID-19 pandemic in Nigeria. The study included 140 participants. Chen et al. [86] studied user satisfaction in the context of using online education platforms in China during COVID-19. The work used a combination of questionnaires and a back propagation neural network.

Drane et al. [87] performed a comprehensive review of existing works to present the impact of 'learning at home' on the educational outcomes of vulnerable children in Australia during the COVID-19 pandemic. The work of Mukuna et al. [88] explored the perceived challenges of online teaching encountered by educators in a school in the Thabo Mofutsanyana District in South Africa. A total of 6 educators participated in this study. In [89], Hsiao presented the results of a study to explore the influences of course type and gender on distance learning performance. A total of 18,085 students from a university in Taiwan comprised the sample size of this study. Nafrees et al. [90] performed an analysis to determine the factors of awareness of students about online learning among undergraduate students at Southeastern University, a university in Sri Lanka. The study comprised about 400 questionnaires, and a total of 310 responses from students were analyzed by the authors. The findings showed that most students preferred to use WebEx over other platforms for their online education due to the user-friendliness of WebEx.

In terms of mining relevant conversations related to a specific topic on Twitter since the outbreak of COVID-19, the prior works in this field have focused on the development of datasets for healthcare misinformation [91], misleading information [92], vaccine misinformation [93], patient identification [94], updates related to vaccine development [95], and rumors related to COVID-19 [96].

Despite these emerging works in the fields of online learning and the development of Twitter datasets, there exist multiple limitations. First, these works in the field of online learning have been confined to studying or analyzing the success or failure, degrees of acceptance, and associated factors related to online learning in specific geographic regions such as Pakistan [61], Indonesia [62,63,66,67], Philippines [64], UAE [65], Saudi Arabia [68], Jordan [69], Kazakhstan [70], Ghana [71], Saudi Arabia [72], Bangladesh [73], United States [74,77], India [75], United Kingdom [76], Germany [77], UAE [78], Ukraine [79], Russia [80], Spain [81], Greece [82], Italy [83], Austria [84], Nigeria [85], China [86], Australia [87], South Africa [88], Taiwan [89], and Sri Lanka [90], and not at a global level. Second, due to the lack of datasets such as Twitter conversations related to online learning from global users, the data that were analyzed in these studies were mostly in the form of surveys that were conducted in these respective geographic regions. Third, the Twitter datasets related to COVID-19 [91-96] do not focus on online learning and the ongoing chatter on Twitter about the same amidst the global rise of COVID-19 cases due to the Omicron variant. The dataset proposed in this paper aims to address all these limitations.

## 3. Methodology

This section describes the methodology that was followed for the development of this dataset publicly available at https://doi.org/10.5281/zenodo.6837118. The dataset contains a total of 52,984 Tweet IDs that correspond to the same number of tweets about online learning, which were publicly posted on Twitter from 9th November 2021 to 13th July 2022. This section also outlines how this work and the associated dataset development is in compliance with the privacy policy, developer agreement, and guidelines for content redistribution of Twitter, as well as follows the FAIR principles (Findability, Accessibility, Interoperability, and Reusability) principles for scientific data management. These are discussed in Sections 3.1, 3.2, and 3.3, respectively.

### 3.1. Process for Dataset Development

As this work focuses on developing a Twitter dataset, the privacy policy, developer agreement, and guidelines for content redistribution of Twitter [97,98] were thoroughly studied, and after studying the same, it was concluded that mining relevant tweets from Twitter to develop a dataset (comprising only Tweet IDs) is in compliance with all these policies of Twitter. Therefore, this dataset contains only Tweet IDs and does not contain any other information related to the respective Tweets that were mined. A detailed explanation of this compliance is mentioned in Section 3.2.

The tweets were collected by using the Search Twitter "operator" [99] available in RapidMiner studio [100] and the Advanced Search feature of the Twitter API. RapidMiner is a data science platform that allows the development, implementation, and testing of various algorithms, processes, and applications in the fields of Big Data, Data Mining, Data Science, Artificial Intelligence, Machine Learning, and their related areas. There are various RapidMiner products available such as RapidMiner Studio, RapidMiner AI Hub, and RapidMiner Radoop. For this work, the RapidMiner studio, version 9.10, was downloaded and installed on a laptop with the Microsoft Windows 10 Home operating system with Intel(R) Pentium(R) Silver N5030 CPU @ 1.10GHz, 1101 Mhz, 4 Core(s), and 4 Logical Processor(s). In the RapidMiner platform, "process" and "operator" are two commonly used terminologies. An "operator" represents a specific function or operation, for instance, fetch data from a social media platform such as Twitter based on a specific set of guidelines or to perform a specific operation on a dataset. RapidMiner has a number of in-built "operators". It also allows users to develop "operators" from scratch. A collection of "operators" that are connected in a logical and executable sequence to achieve a desired purpose is called a "process". A "process" may also contain just one "operator" if the complete functionality of the "process" can be found in one in-built or user-defined "operator". The Search Twitter "operator", an in-built "operator" of RapidMiner, works by connecting with the Twitter API and by complying with the Twitter API standard search policies [101] to fetch tweets between two given dates that contain one or more keywords or phrases which are provided as input to this "operator". As there are different keywords that Twitter users can use to refer to both COVID-19, the Omicron variant, and online learning, therefore a bag of words was developed based on studying commonly used synonyms, phrases, and terms used to refer to online learning [102], COVID-19 and the Omicron variant [103]. These synonyms, terms, and phrases, all of which were included in the data collection process, are shown in Table 1.

There are various forms of educational structures and educational systems followed by different countries all over the world. For instance, in the United States, early childhood education is followed by primary school (also called elementary school), middle school, secondary school (also called high school), and then postsecondary (tertiary) education. Postsecondary education includes non-degree programs that lead to certificates and diplomas plus six-degree levels: associate, bachelor, first professional, master, advanced intermediate, and research doctorate. The US system does not offer a second or higher doctorate but does offer post-doctorate research programs [104]. A different educational structure is followed in India [105]. The school system in India has four levels: lower primary school (age 6 to 10), upper primary school (age 11 and 12), high school (age 13 to 15), and higher secondary school (age 17 and 18). The lower primary school is divided into five "standards", upper primary school into two, high school into three, and higher secondary school into two. Another different educational structure can be seen in the United Kingdom (UK). The education system in the UK is divided into four main parts, primary education, secondary education, further education, and higher education. Children in the UK have to legally attend primary and secondary education, which runs from about five years old until the student is 16 years old. The education system in the UK is also split into "key stages" - Key Stage 1 (age 5 to 7), Key Stage 2 (age 7 to 11), Key Stage 3 (age 11 to 14), Key Stage 4 (age 14 to 16) [106]. This study focuses on collecting tweets about online education or online learning on a global scale (and not tweets originating from any specific country specific to its educational structure or educational system). So, a comprehensive list of keywords (as shown in Table 1) was developed that would most commonly be used to refer to online education or online learning in different parts of the world irrespective of the educational structure followed in that specific geographic region. The effectiveness of this approach can be seen from the different worldwide educational systems that are the subject matters of the tweets present in the dataset proposed as a result of this work. For instance, in this dataset, Tweet ID: 1458685065152450565 refers to online education in India, Tweet ID: 1462489169079513090 refers to online education in the United States, Tweet ID: 1462475208644874242 refers to online education in Pakistan, Tweet ID: 1462373712389238787 refers to online education in Indonesia, Tweet ID: refers to online education in the UK, Tweet ID: 1462357217479434241 refers to online education in Ukraine, Tweet ID: 1462512737402109952 refers to online education in Nigeria, Tweet ID: 1462315144411856897 refers to online education in Spain, Tweet ID: 1462411445035941891 refers to online education in Malaysia, and so on.

**Table 1.** List of commonly used synonyms, terms, and phrases for online learning and COVID-19

| Terminology | List of synonyms and Terms |
|---|---|
| COVID-19 | Omicron, COVID, COVID19, coronavirus, coronaviruspandemic, COVID-19, corona, coronaoutbreak, omicron variant, SARS CoV-2, corona virus |
| online learning | online education, online learning, remote education, remote learning, e-learning, elearning, distance learning, distance education, virtual learning, virtual education, online teaching, remote teaching, virtual teaching, online class, online classes, remote class, remote classes, distance class, distance classes, virtual class, virtual classes, online course, online courses, remote course, remote courses, distance course, distance courses, virtual course, virtual courses, online school, virtual school, remote school, online college, online university, virtual college, virtual university, remote college, remote university, online lecture, virtual lecture, remote lecture, online lectures, virtual lectures, remote lectures |

Tweets were searched using this "process" that comprised the Search Twitter "operator" in a way that it consisted of at least one synonym or term or phrase used to refer to COVID-19 and at least one synonym or term or phrase used to refer to online learning. The Search Twitter "operator" is not case-sensitive, so it returned the tweets based on keyword matching by ignoring the case (uppercase or lowercase).

The output of this RapidMiner "process" comprised of multiple attributes such as the Tweet ID, Tweet Source (the source used to post the Tweet such as Twitter for Android, Twitter for IOS, etc.), Text of the Tweet, Retweet count, and the username of the Twitter user who posted the Tweet, all of which is public information that can be mined in compliance with the guidelines set forth in the Twitter API standard search policies. However, as per the developer policy, privacy policy, and content redistribution guidelines of Twitter, all the attributes other than the Tweet IDs were deleted by using data filters. Therefore, the dataset consists of only Tweet IDs. These Tweet IDs were grouped into different .txt files based on the timeline of the associated tweets. The description and details of these dataset files are presented in Section 4.

The complete information associated with a tweet, such as the text of a tweet, user name, user ID, timestamp, retweet count, etc., can be obtained from a Tweet ID by following a process known as hydration of Tweet ID [107]. Researchers in the field of Big Data, Data Mining, and Natural Language Processing, with a specific focus on Twitter research, have developed multiple tools for hydration of Tweet IDs. Some of the most commonly used tools include – the Hydrator app [108], Social Media Mining Toolkit [109], and Twarc [110], all of which work by complying with the policies of accessing the Twitter API. Any of these tools can be used on this dataset to obtain the associated information, such as the text of a tweet, user name, user ID, timestamp, and retweet count for all the Tweet IDs. A step-by-step process on how to use one of these tools – the Hydrator app for hydrating all the Tweet IDs in this dataset is mentioned in Appendix A.

A couple of things are worth mentioning here. First, Twitter allows users the option to delete a tweet which would mean that there would be no retrievable Tweet text and other related information (upon hydration) for a Tweet ID of a deleted tweet. All the Tweet IDs available in this dataset correspond to tweets that have not been deleted at the time of writing this paper. Second, the Twitter API's search feature does not return an exhaustive list of tweets that were posted in a specific date range. So, it is possible that multiple tweets that might have been posted in between this date range were not returned by the Twitter API's search feature when the data collection was performed and are thus not a part of this dataset.

### 3.2. Compliance with Twitter Policies

The privacy policy of Twitter [97] states – *"Twitter is public and Tweets are immediately viewable and searchable by anyone around the world"*. To add, the Twitter developer agreement [98] defines tweets as *"public data"*. The guidelines for Twitter content redistribution [98] state – *"If you provide Twitter Content to third parties, including downloadable datasets or via an API, you may only distribute Tweet IDs, Direct Message IDs, and/or User IDs (except as described below). It also states - "We also grant special permissions to academic researchers sharing Tweet IDs and User IDs for non-commercial research purposes. Academic researchers are permitted to distribute an unlimited number of Tweet IDs and/or User IDs if they are doing so on behalf of an academic institution and for the sole purpose of non-commercial research."* Therefore, it may be concluded that mining relevant tweets from Twitter to develop a dataset (comprising only Tweet IDs) and to share the same is in compliance with the privacy policy, developer agreement, and content redistribution guidelines of Twitter.

### 3.3. Compliance with FAIR

This section outlines how this dataset is compliant with the FAIR principles (Findability, Accessibility, Interoperability, and Reusability) principles for scientific data management [111]. The dataset is findable as it has a unique and permanent DOI, which was assigned by Zenodo. The dataset is accessible online. It is interoperable due to the use of .txt files for data representation that can be downloaded, read, and analyzed across different computer systems and applications. The dataset is re-usable as the associated tweets and related information, such as user ID, user name, retweet count, etc., for all the Tweet IDs can be obtained by the process of hydration in compliance with Twitter policies (Appendix A) for data analysis and interpretation.

## 4. Data Description

This section provides a detailed description of this dataset. The raw version of the dataset comprised 67,319 tweets. This included multiple duplicate tweets. The duplicate tweets were recorded mostly because several Twitter users used a list of different hashtags referring to either online learning and/or the Omicron variant of COVID-19 in the same tweet, probably for increased audience engagement. For instance, as per the methodology described in Section 3, Tweet ID: 1464533235367510019 was captured twice as it contains two synonyms ("omicron" and "covid") from the list of synonyms presented in Table 1. Therefore, after the data collection process was completed as described in Section 3, data preprocessing and data cleaning were performed using RapidMiner to remove duplicate tweets. After the removal of duplicate tweets, the dataset comprised 52,984 Tweet IDs corresponding to the same number of tweets about online learning posted on Twitter between 9th November 2021 (the sample collected on this date was the first case of Omicron) to 13th July 2022 (the most recent date at the time of re-submission of this paper to this journal after the completion of the first round of peer review and the subsequent editorial decision). The dataset is available at https://doi.org/10.5281/zenodo.6837118. The dataset comprises nine .txt files. Table 2 presents the description of each of these dataset files along with the number of Tweet IDs present in each of them. As can be seen from Table 2, the most number of Tweets were posted in January 2022. The fact that the tweets of only 13 days in July 2022 were mined is the likely reason why July 2022 accounts for the least number of tweets as per this table.

**Table 2.** Description of all the files present in this dataset that comprises tweets about online learning during the current COVID-19 Omicron Wave.

| Filename | No. of Tweet IDs | Date Range of the associated tweets |
|---|---|---|
| TweetIDs_November_2021.txt | 1283 | November 1, 2021 to November 30, 2021 |
| TweetIDs_December_2021.txt | 10545 | December 1, 2021 to December 31, 2021 |
| TweetIDs_January_2022.txt | 23078 | January 1, 2022 to January 31, 2022 |
| TweetIDs_February_2022.txt | 4751 | February 1, 2022 to February 28, 2022 |
| TweetIDs_March_2022.txt | 3434 | March 1, 2022 to March 31, 2022 |
| TweetIDs_April_2022.txt | 3355 | April 1, 2022 to April 30, 2022 |
| TweetIDs_May_2022.txt | 3120 | May 1, 2022 to May 31, 2022 |
| TweetIDs_June_2022.txt | 2361 | June 1, 2022 to June 30, 2022 |
| TweetIDs_July_2022.txt | 1057 | June 1, 2022 to July 13, 2022 |

Table 3 presents some characteristic features of this dataset. As can be seen from Table 3, the tweets are present in 38 different languages in this dataset. The most common language is English (50539 Tweets), which is followed by Indonesian (527 Tweets), Tagalog (525 Tweets), Estonian (364 Tweets), Spanish (236 Tweets), Hindi (179 Tweets), and 33 other languages. All these tweets were posted on 237 different days between November 9, 2021, and July 13, 2022. The highest number of Tweets was recorded on January 5, 2022 (2067 Tweets), which is followed by January 6, 2022 (1592 Tweets), January 3, 2022 (1465 Tweets), January 4, 2022 (1355 Tweets), and the other dates. A total of 17950 distinct Twitter users posted these tweets, who have a total follower favorite count of 4345192697. The combined follower favorite count and retweet count of all the tweets present in this dataset are 3273263 and 556980, respectively. A total of 5722 Tweets present in this dataset were posted by Twitter users with a verified Twitter account, and the remaining Tweets came from an unverified Twitter account. The number of distinct URLs that can be found embedded in these Tweets is 7869. The URL that occurs the most number of times (30 times) in the tweets points to a list of online courses for COVID-19 safety at work [112]. The URL that occurs the second most number of times (29 times) is a YouTube video that is also an online course on COVID-19 [113].

**Table 3.** Characteristic Features of this dataset that comprises tweets about online learning during the current COVID-19 Omicron Wave.

| Characteristic Feature | Count |
|---|---|
| Languages in which the Tweets are available | 38 |
| Distinct days when the Tweets were posted | 237 |
| Distinct users who posted the Tweets | 17950 |

| Total followers count of all the Twitter users who posted the Tweets | 4345192697 |
| Number of Tweets from a verified Twitter account | 5722 |
| Number of Tweets from an unverified Twitter account | 47262 |
| Total favorite count of all the Tweets | 3273263 |
| Total retweet count of all the Tweets | 556980 |
| Distinct URLs embedded in the Tweets | 7869 |

## 5. Potential Applications: Brief Overview

This dataset of more than 50,000 Tweet IDs is expected to help advance interdisciplinary research in different fields such as Big Data, Data Science, Data Mining, Natural Language Processing, Healthcare, and their related disciplines. A few potential applications and use-case scenarios that may be investigated using this dataset include performing sentiment analysis [114], performing aspect-based sentiment analysis [115], predicting popular tweets [116], detecting sarcasm [117], developing topic modeling [118], tracking retweeting patterns [119], ranking tweets [120], performing content value analysis [121], tracking credibility of information [122], detecting conspiracy theories [123], predicting emoji usage patterns [124], studying the relevance of information [125], detecting satire [126], detecting deception [127], extracting categorical topics and emerging issues [128], characterizing Twitter users [129], and detection of Twitter user demographics [130] in the context of Twitter chatter related to online learning during the current Omicron wave of COVID-19.

## 6. Conclusion

The outbreak of COVID-19 led to schools, colleges, and universities in almost all parts of the world closing and transitioning to online learning. The development of vaccines and other forms of treatment towards the end of 2020 led to some of these educational institutions re-opening and starting to function in a hybrid as well as in a completely in-person manner. The recent surge of COVID-19 cases globally due to the Omicron variant, the most immune evasive variant of COVID-19 that presents very strong resistance against antibody-based or plasma-based treatments, has resulted in several such educational institutions switching to online learning once again. This has led to an increase in the number of online conversations, specifically on Twitter, related to online learning since the first detected case of the Omicron variant in November 2021. Mining such tweets to develop a dataset would serve as a data resource for interdisciplinary research related to the analysis of interest, views, opinions, perspectives, attitudes, and feedback towards online learning during the current surge of COVID-19 cases caused due to this variant. The prior works in this field did not focus on the development of a similar data resource. Therefore, this work presents an open-access dataset of more than 50,000 Tweet IDs (that correspond to the same number of tweets) about online learning posted on Twitter between 9th November 2021 (the sample collected on this date was the first case of Omicron) to 13th July 2022 (the most recent date at the time of re-submission of this journal paper after the completion of the first round of peer review and the subsequent editorial decision). It contains 52,984 Tweet IDs that correspond to the same number of tweets about online learning posted during the ongoing COVID-19 Omicron wave. The dataset is compliant with the privacy policy, developer agreement, and guidelines for content redistribution of Twitter, as well as with the FAIR principles (Findability, Accessibility, Interoperability, and Reusability) principles for scientific data management. The paper also briefly outlines a few potential research directions that may be investigated using this dataset. Future work on this project would involve updating the dataset with more recent tweets to ensure that the scientific community has access to the recent data in this regard.

**Supplementary Materials:** Not applicable.

**Funding:** This research received no external funding

**Institutional Review Board Statement:** Not applicable

**Informed Consent Statement:** Not applicable.

**Data Availability Statement:** The data presented in this study are publicly available at https://doi.org/10.5281/zenodo.6837118

**Conflicts of Interest:** The author declares no conflict of interest

## Appendix A

The following is the step-by-step process for using the Hydrator app [105] to hydrate this dataset or, in other words, to obtain the text of the tweet, user ID, user name, retweet count, language, tweet URL, source, and other public information related to all the Tweet IDs present in this dataset. The Hydrator app works in compliance with the policies for accessing and calling the Twitter API.

1. Download and install the desktop version of the Hydrator app from https://github.com/DocNow/hydrator/releases.
2. Click on the "Link Twitter Account" button on the Hydrator app to connect the app to an active Twitter account.
3. Click on the "Add" button to upload one of the dataset files (in .txt format, such as TweetIDs_Omicron.txt). This process adds the dataset file to the Hydrator app.
4. If the file upload is successful, the Hydrator app will show the total number of Tweet IDs present in the file. For instance, for the file - "TweetIDs_June_2022.txt ", the app would show the Number of Tweet IDs as 2361.
5. Provide details for the respective fields: Title, Creator, Publisher, and URL in the app, and click on "Add Dataset" to add this dataset to the app.
6. The app would automatically redirect to the "Datasets" tab. Click on the "Start" button to start hydrating the Tweet IDs. During the hydration process, the progress indicator would increase, indicating the number of Tweet IDs that have been successfully hydrated and the number of Tweet IDs that are pending hydration.
7. After the hydration process ends, a .jsonl file would be generated by the app that the user can choose to save on the local storage.
8. The app would also display a "CSV" button in place of the "Start" button. Clicking on this "CSV" button would generate a .csv file with detailed information about the tweets, which would include the text of the tweet, user ID, user name, retweet count, language, tweet URL, source, and other public information related to the tweet.
9. Repeat steps 3-8 for hydrating all the files of this dataset.